\documentclass[12pt]{article} 

\usepackage{scicite}
\usepackage{graphicx}
\usepackage{times}
\usepackage{amssymb}
\usepackage{amsmath}
\usepackage{color}
\usepackage[ruled]{algorithm2e}
\usepackage{caption}
\usepackage{braket}
\usepackage[normalem]{ulem}
\usepackage{placeins}  

\usepackage[colorlinks,
linkcolor=blue,       
anchorcolor=blue,  
citecolor=blue,        
urlcolor=blue,hyperindex]{hyperref}

\newenvironment{sciabstract}{\begin{quote} \bf}
{\end{quote}}

\begin{document} 

\title{{Adaptive cold-atom magnetometry mitigating the trade-off between sensitivity and dynamic range}}

\author{Zhu Ma$^{1,2\dag}$, Chengyin Han$^{1,3\dag}$, Zhi Tan$^{1}$, Haihua He$^{1}$, Shenzhen Shi$^{1}$, \\
Xin Kang$^{1}$, Jiatao Wu$^{1,2}$, Jiahao Huang$^{1,2}$, Bo Lu$^{1,3\ast}$, Chaohong Lee$^{1,3\ast}$\\
\\
\normalsize{$^{1}$Institute of Quantum Precision Measurement,}\\
\normalsize{State Key Laboratory of Radio Frequency Heterogeneous Integration,}\\
\normalsize{College of Physics and Optoelectronic Engineering, Shenzhen University, Shenzhen 518060, China.}\\
\normalsize{$^{2}$Laboratory of Quantum Engineering and Quantum Metrology, School of Physics and Astronomy,}\\ 
\normalsize{Sun Yat-Sen University (Zhuhai Campus), Zhuhai 519082, China.}\\
\normalsize{$^{3}$Quantum Science Center of Guangdong-Hong Kong-Macao Greater Bay Area (Guangdong),}\\
\normalsize{Shenzhen 518045, China.}\\
\\
\normalsize{$^\dag$These authors contributed equally to this work.}\\
\normalsize{$^\ast$To whom correspondence should be addressed;}\\ 
\normalsize{E-mail: lubo1982@szu.edu.cn, chleecn@szu.edu.cn.} 
}
\date{}
\maketitle

\baselineskip24pt

\section*{Abstract}
\begin{sciabstract}
Cold-atom magnetometers can achieve an exceptional combination of superior sensitivity and high spatial resolution. 
One key challenge these quantum sensors face is improving the sensitivity within a given timeframe while preserving a high dynamic range.
Here, we experimentally demonstrate an adaptive entanglement-free cold-atom magnetometry with both superior sensitivity and high dynamic range.
Employing a tailored adaptive Bayesian quantum estimation algorithm designed for Ramsey interferometry using coherent population trapping (CPT), cold-atom magnetometry facilitates adaptive high-precision detection of a direct-current (d.c.) magnetic field with high dynamic range.
Through implementing a sequence of correlated CPT-Ramsey interferometry, the sensitivity significantly surpasses the standard quantum limit with respect to total interrogation time.
We yield a sensitivity of 6.8$\pm$0.1 pico
tesla per square root of hertz over a range of 145.6 nanotesla, exceeding the conventional frequentist protocol by 3.3$\pm$0.1 decibels.
Our study opens avenues for the next generation of adaptive cold-atom quantum sensors, wherein real-time measurement history is leveraged to improve their performance.

\end{sciabstract}

\section*{Teaser}
Cold-atom magnetometry achieves sensitivity surpassing the optimal frequentist level while maintaining a high dynamic range.

\section*{Introduction}

Atomic magnetometers, known for their ultrahigh sensitivity, user-friendly operation, and compact design, have been utilized in a wide range of fields, from fundamental research \cite{Aslam_2017,Safronova_2018,Hu_2020,Smorra_2019,Jiang_2021,Su_2021,Wang_2023} to practical applications \cite{Matti_1993,Boto_2018,Hill_2019,Zhang_2020}. 
Extreme sensitivity (less than femtotesla per square root of hertz) can generally be achieved using large atomic ensembles, such as thermal atomic samples in vapor cells \cite{Dang_2010}.
However, these setups are limited by their inherently low spatial resolution, typically at effective linear dimensions of several millimeters \cite{Mitchell_2020}, making them unsuitable for high-spatial-resolution magnetic field sensing. 
In contrast, cold-atom magnetometers offer superior spatial resolution and long coherence time, making them particularly appealing for precise magnetometry applications \cite{Fregosi_2020,Cohen_2019,Vengalattore_2007,Eto_2013,Jasperse_2017,Sekiguchi_2021}.

However, cold-atom magnetometers encounter a great challenge in overcoming the trade-off between sensitivity and dynamic range.  
To detect a d.c. magnetic field, these quantum sensors generally operate based upon the Ramsey interferometry of two magnetic-sensitive states. 
Due to the d.c. magnetic field $B$, a Zeeman shift $f_B = \Delta m_F\gamma B $ appears between the two states, where \(\gamma\) is the gyromagnetic ratio and $\Delta m_F$ is the difference of magnetic quantum numbers.
In a Ramsey interferometry, the first $\pi/2$ pulse prepares a superposition state, which will accumulate a phase \(\phi = 2\pi f_B T_R\) from the magnetic field $B$ during an interrogation time \(T_R\), and the second $\pi/2$ pulse transforms the information of $\phi$ into the final population. 
The measurement uncertainty $\Delta B$ generally obeys the standard quantum limit (SQL), which scales as $\Delta B \propto {1}/({2\pi |\Delta m_F\gamma|  \sqrt{T_R T N}})$ with respect to the total interrogation time $T$ (which is the sum of interrogation times across all experimental cycles) and the total particle number $N$.
That is, a long interrogation time $T_R$ corresponds to a high sensitivity.
However, due to phase ambiguities~\cite{Said_2011,Waldherr_2011,Nusran_2012}, long interrogation time $T_R$ will reduce the dynamic range \(B_{\rm max} = 1/(2 |\Delta m_F\gamma| T_R)\).
High-spatial-resolution magnetometry with Bose condensed atoms has achieved a high sensitivity of 5.0~pT$/\sqrt{\rm Hz}$, 
but the corresponding dynamic range is very low ($B_{\rm max} = 1/(2 |\Delta m_F\gamma| T_R) = 238.1~\textrm{pT}$  with $T_R = 300~\textrm{ms}$)~\cite{Sekiguchi_2021}.
Up to date, it remains a dilemma to improve the dynamic range of cold-atom magnetometry without compromising its sensitivity.

In addition to the correlations between particles, the correlations between interrogation times can be utilized to enhance the sensitivity of cold-atom magnetometry. 
In quantum metrology, multiparticle quantum entanglement (a typical quantum correlation) has been extensively used to improve the sensitivity with respect to the total particle number $N$ from SQL ($\propto N^{-0.5}$) to sub-SQL scaling ($\propto N^{-\alpha}$ with $0.5<\alpha\leq 1$)~\cite{Gross_2010,Muessel_2014,Ockeloen_2013,Sewell_2012,Mao_2023,Huang_2024}. 
Although multiparticle quantum entanglement may improve sensitivity scaling, the challenge of preparing large-particle-number entangled states and the fragility of those entanglement in realistic environments limits the attainable measurement precision, often preventing it from exceeding that of entanglement-free systems \cite{Huang_2024}.
Alternatively, Bayesian quantum parameter estimation, the cooperation of quantum parameter estimation and Bayesian statistics, offers a unique opportunity to improve the sensitivity with respect to the total interrogation time $T$ from {SQL} ($\propto T^{-0.5}$) to sub-{SQL} scaling ($\propto T^{-\alpha}$ with $0.5<\alpha\leq 1$) \cite{Herbschleb_2021,Higgins_2007,Wiebe_2016,Cappellaro_2012,Ferrie_2012,Nusran_2012,Waldherr_2011,Santagati_2019,Bonato_2016,Turner_2022}. 
In Bayesian quantum parameter estimation with single-particle systems, the likelihood function is usually a periodic cosine function and the measurement outcome is binary data.
Using an ensemble of identical atoms, one can yield a signal-to-noise ratio (SNR) that is higher than that given by single-particle systems. 
However, it requires tailoring and updating the likelihood function and posterior probability based on the signals provided by an ensemble of atoms rather than a single-particle system.
Up to now, it has not yet been demonstrated how to enhance cold-atom magnetometry through Bayesian quantum parameter estimation.

In this article, we present adaptive measurements of d.c. magnetic fields that achieve both high sensitivity and high dynamic range, utilizing a cold-atom magnetometer based on $^{87}$Rb atoms in coherent population trapping (CPT).
Unlike conventional frequentist measurements, whose interrogation times are fixed, our adaptive Bayesian quantum estimation utilizes a sequence of correlated CPT-Ramsey interferometry with exponentially increasing interrogation times and adaptively updated auxiliary phases.
We experimentally demonstrate that the measurement sensitivity with respect to the total interrogation time significantly exceeds the standard quantum limit.
Consequently, our Bayesian cold-atom CPT magnetometer achieves a sensitivity of \(6.8\pm0.1 \ \mathrm{pT}/\sqrt{\mathrm{Hz}}\) with a dynamic range of 145.6~nT.
This represents a $3.3\pm0.1$~dB improvement in sensitivity and 14.6~dB increase in dynamic range compared to the best sensitivity of \(14.7\pm 0.4 \ \mathrm{pT}/\sqrt{\mathrm{Hz}}\) and a dynamic range of 5.0~nT in the corresponding frequentist protocol using $T_R=T_{\rm max}$.
Bayesian quantum estimation leverages real-time measurement history to achieve both high dynamic range and high sensitivity, enabling the next generation of adaptive cold-atom quantum sensors.

\section*{Results}
\subsection*{Experimental setup}

\begin{figure*}[htp]
\centering
\includegraphics[width=1\linewidth]{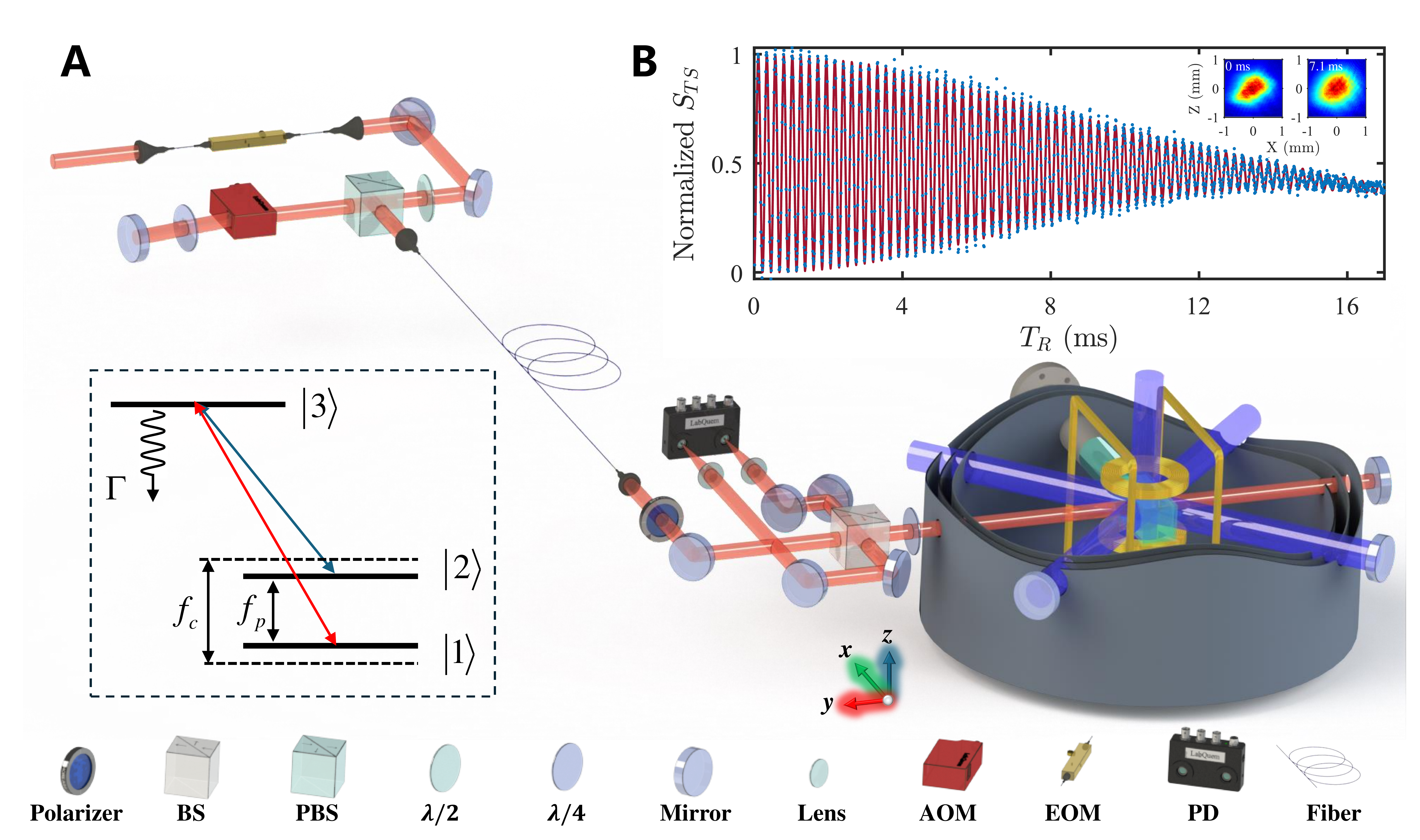}
\caption{
\textbf{Experimental setup of cold-atom CPT magnetometry.}
\textbf{(A)} The CPT beam is generated and manipulated by an EOM, an AOM, and a polarizer.
To enhance the robustness of the signal against power noise, the CPT beam is separated into transmitted and reflected beams using a BS with a 70:30 split ratio. 
The $\sigma^- - \sigma^-$ configuration of the CPT beam is achieved by employing a quarter-wave plate and the CPT beam is aligned with the applied bias magnetic field.
The inset shows a typical three-level $\Lambda$ system of $^{87}$Rb used in our experiment. 
The two magnetic-sensitive states
$(\ket{1}=\ket{F = 1, m_F = -1}$ and $\ket{2}=\ket{F = 2, m_F = -1})$ are coupled through the excited state $(\ket{3}=\ket{F^{\prime} = 2, m_F = -2})$.
AOM: acousto-optic modulator, EOM: electro-optic modulator, BS: beam splitter, PBS: polarization beam splitter, PD: photodetector, $\lambda /$2: half-wave plate, $\lambda /$4: quarter-wave plate.
\textbf{(B)} Time-domain CPT-Ramsey fringes as a function of $T_R$ under fixed $B$ and detuning $\Delta f$. The red line represents the fitting of the fringes to obtain the coherence time {$T_\chi\approx10.0$} ms (see Materials and Methods for details). Inset: The absorption imaging of atomic cloud at release times of 0 ms and {7.1} ms. 
We determine the spatial resolution by measuring the space occupied by the falling process of the atom cloud, which is approximated as a cylinder.
}
\label{Figure1}
\end{figure*}

We combine cold atoms and CPT to implement spatially resolved magnetometry via CPT-Ramsey interferometry~\cite{Han_2024}.
We use a bichromatic light field to couple the ground-state Zeeman levels \(\vert F=1, m_F=-1\rangle\) and \(\vert F=2, m_F=-1\rangle\) to the excited state \(\vert F'=2, m_F=-2\rangle\) in the D1 line of \(^{87}\)Rb (inset of Fig. \ref{Figure1}A).
Based on our first-generation experimental apparatus \cite{Fang_2021,Han_2021,Li_2024,Zhan_2024}, we build a more compact experimental apparatus with magnetic shielding, as shown in Fig. \ref{Figure1}A.
$^{87}$Rb atoms are initially captured in a magneto-optical trap (MOT) for 50 ms and are then further cooled using polarization gradient cooling (PGC). 
To suppress the influence of magnetic field relaxation caused by the MOT, we reduce the coil current in segments and implement magnetic field relaxation within 10 ms (see details in Supplementary Material Section 1).
Consequently, the PGC is achieved in this 10 ms period.
After PGC, we obtain about $3 \times 10^7$ cold atoms with a temperature around {13} $\mu$K.
The atoms then fall freely and are interrogated by the left-circularly polarized CPT beams aligned with the bias magnetic field $B$. 
The bichromatic light field is generated from a laser modulated by a fiber-coupled electro-optic modulator (EOM), which is driven by a microwave (MW) synthesizer with a frequency \( f \) approximately equal to the \(^{87}\)Rb ground-state hyperfine splitting frequency.
An acousto-optic modulator (AOM) is used to generate the CPT beam pulses.

We use a timing diagram consisting of a 300-$\mu$s CPT preparation pulse $\tau_p$, followed by an interrogation time $T_R$ and a 50-$\mu$s detection pulse $\tau_d$ for each experimental cycle.
The CPT beam is separated into transmitted and reflected beams by a beam splitter with a 70:30 ratio. 
The reflected beam is detected by one receiver of the balanced photodetector (PD) as $S_N$.
The transmission beam passes through a quarter-wave plate and is converted into left-handed circularly polarized light. 
Then this circularly polarized light interrogates the cold atoms by propagating through the vacuum chamber and reflecting back by a mirror.
The spacing between the retroreflecting mirror and the atoms is an integer multiple of the half-wavelength of the MW to keep the CPT signal amplitude maximal. 
The beam reaches the beam splitter again and is reflected in another receiver of the balanced PD as $S_T$.
The corresponding signals ($S_{TS}$) are proportional to the difference in photocurrent between two receivers of the balanced PD, which can reduce the effect of intensity noise on the CPT-Ramsey signals.

We perform the CPT-Ramsey interferometry in time domain to acquire the coherence time $T_\chi$ between these two ground-state Zeeman levels.
By setting the detuning $\Delta f = f-f_c$ = -95300~Hz, the corresponding CPT-Ramsey fringes are shown in Fig. \ref{Figure1}B, indicating a Gaussian coherence decay as $T_\chi$ = 10.0~ms (see Materials and Methods for details).
Here, $f_c$ is the clock transition frequency of $\vert F=1, m_F=0\rangle \rightarrow \vert F=2, m_F=0\rangle$ and $f_c\approx6.83$~GHz for \(^{87}\)Rb.
In our magnetometry experiment, the maximal interrogation time $T_R=7.1$ ms corresponds to the optimal sensitivity of the conventional frequentist protocol (see details in Supplementary Material Section
2). 
Within an interrogation time of 7.1~ms, the free-fall distance of the atomic cloud is 0.24~mm.
The corresponding radius of the atomic clouds at the release time of 0 ms and 7.1~ms after PGC are 0.40~mm and 0.47~mm, respectively (see the inset of Fig.~\ref{Figure1}B).
We determined that the spatial resolution of our cold-atom CPT magnetometry is nearly 0.77~mm$^{3}$, based on the space occupied by the falling process of the atom cloud. 

\subsection*{Conventional cold-atom CPT magnetometry}

Under weak magnetic fields, the transition frequency between the two ground-state Zeeman levels can be written as $f_p=f_c+f_B$.
We directly measure $f_p$ by stabilizing the MW synthesizer frequency $f$ to the central fringe of CPT-Ramsey interference in the frequency domain.
This is achieved by alternately probing the sides of central CPT-Ramsey fringe via modulating the MW synthesizer frequency.
The frequency of the MW synthesizer is alternated between the values of $f_p-1/4T_R$ and $f_p+1/4T_R$ from cycle to cycle, where $1/2T_R$ is the width of the central Ramsey fringe.
The magnetic field is then acquired by the relationship of $B = (f_p - f_c) / (\Delta m_F\gamma)$, where $\Delta m_F=-2$ and the $^{87}$Rb gyromagnetic ratio $\gamma\approx7$ Hz/nT.

In order to obtain $f_p$, two CPT-Ramsey interferometry measurements are carried out.
Hence, the averaging time $\tau=2M T_c$ with $M$ independent measurements of $f_p$.
The sensitivity of frequentist measurement $\overline{\eta}_{\rm FMM}$ for averaging time $\tau$ is given as (see Materials and Methods for details), 
\begin{equation}
        {\overline{\eta}_{\rm FMM}=\frac{\sqrt{T_c}}{\pi|\Delta m_F\gamma| T_R \sqrt{N_{\rm eff}}}\propto \frac{B_{\rm max}}{\sqrt{N_{\rm eff}}}.}
\label{eq_sensitivity_averaging_time}
\end{equation}
Here, $N_{\rm eff}$ is the effective particle number determined by the SNR, $T_c = T_R+T_d$ is a fixed cycle period in our experiment, with $T_d$ the dead time needed to prepare, initialize and readout the quantum states.
In our experiment, each CPT-Ramsey cycle takes $T_c =73~\textrm{ms}$, the effective particle number decreases according to $N_{\rm eff}=A e^{-2(T_R/T_{\chi})^2}$ with $A=10755\pm2101$, and we obtain an optimal sensitivity $\overline{\eta}_{\rm FMM}=14.7\pm0.4~\textrm{pT}/\sqrt{\rm Hz}$ with $T_R=7.1~\textrm{ms}$ (see details in Supplementary Material Section
2).
According to Eq. \ref{eq_sensitivity_averaging_time}, one may choose the optimal $T_R$ to achieve the highest sensitivity.
However, the corresponding dynamic range \(B_{\rm max}\) would become very small due to phase ambiguities.
Then a trade-off should be balanced between dynamic range and sensitivity \cite{Yankelev_2020}.
Ignoring the dead time, i.e. $T_d=0$, the sensitivity $\eta_{\rm FMM}$ with respect to the total interrogation time $T=2MT_R$ {(the sum of interrogation times across all the measurement cycles)} can be given as
\begin{equation}
        \eta_{\rm FMM}=\frac{1}{\pi|\Delta m_F\gamma| \sqrt{T_R} \sqrt{N_{\rm eff}}}.
\label{eq_sensitivity_total_phase_accumulation_time}
\end{equation}
Obviously, the sensitivity $\eta_{\rm FMM}$ with respect to the total interrogation time is independent of $T$.
We show this scaling for $T_R=0.245$ ms and $T_R=7.1$ ms (see Fig. \ref{Figure2}C). 
The experimental results are well consistent with this scaling for a short total interrogation time, but the low-frequency noise deteriorates sensitivity when the total interrogation time increases (see the noise spectrum density in Supplementary Material Section 3).
The sensitivity versus the total interrogation time $T$ is given by $\eta_{\rm FMM} = \Delta B_{\rm FMM} \sqrt{T}$ \cite{Waldherr_2011,Nusran_2012,Santagati_2019}, which implies that the uncertainty of frequentist measurements can be expressed as
\begin{equation}
        \Delta B_{\rm FMM}=\frac{1}{\pi|\Delta m_F\gamma| \sqrt{N_{\rm eff}}\sqrt{T_RT}}.
\label{eq_sandard_deviation_ramsey_time}
\end{equation}
It suggests that the uncertainty versus the total interrogation time obeys the SQL: $\Delta B_{\rm FMM}\propto~T^{-0.5}$, as shown in Fig.~\ref{Figure2}B.

\subsection*{Bayesian cold-atom CPT magnetometry}
To achieve high sensitivity without sacrificing dynamic range, we develop an adaptive Bayesian cold-atom CPT magnetometry.
Unlike frequentist measurements, we use a sequence of correlated phase-domain CPT-Ramsey interferometry to implement Bayesian quantum estimation.
As the shape and period of the phase-domain CPT-Ramsey fringes are invariant for different interrogation times, we can normalize the interference signal to reduce the influence of contrast changes caused by decoherence.
Normalization is implemented by preliminary measurement of the maximum $S_{TS}^{\rm max}(T_R)$ and minimum $S_{TS}^{\rm min}(T_R)$ of Ramsey fringes in the phase domain.
We obtain the normalized signal $p = \left(S_{TS}(T_R) - S_{TS}^{\rm min}(T_R)\right)/\left(S_{TS}^{\rm max}(T_R) - S_{TS}^{\rm min}(T_R)\right)$.
The normalized phase-domain Ramsey signals of the atoms that occupy the magnetic sensitive state $\ket{F=2,m_F=-1}$ with respect to $\phi_c$ can be given as $p_e=\frac{1}{2}\{1-\cos{[2\pi(\Delta f-f_B)T_R+\phi_c]}\}$ \cite{Hemmer_1989,Pati_2015,Fang_2023}.
Here, $\phi_c$ is an auxiliary phase controlled by adjusting the phase difference between the two pulses of the CPT-Ramsey sequence (see details in Supplementary Material Section 4).

\begin{figure*}[tp]
	\includegraphics[width=\linewidth]{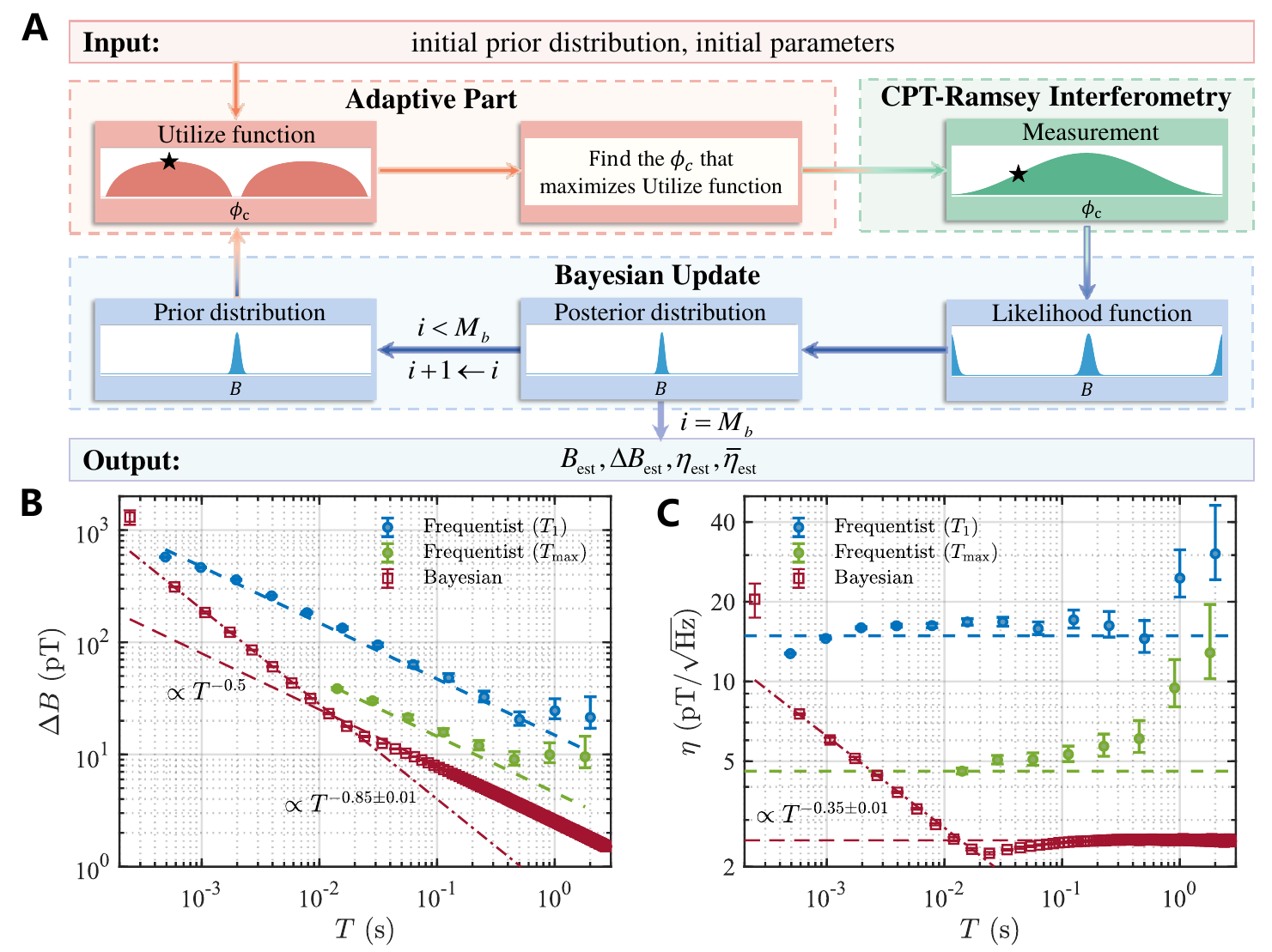}
	\caption{
\textbf{Adaptive Bayesian cold-atom CPT magnetometry.}
\textbf{(A)} Schematic of adaptive Bayesian cold-atom CPT magnetometry. 
The initial prior distribution can be given as a uniform distribution.
The optimal auxiliary phase $\phi_c$ that maximizes the Utilize function (see Eq. \ref{ee:utilize}) can be obtained adaptively.
The likelihood function is obtained by measuring population probability with $i$-th interrogation time and optimal auxiliary phase $\phi_c$ in phase-domain CPT-Ramsey interferometry.
The posterior distribution is updated through Bayes’ formula and then replaces the prior function in each subsequent update until $i=M_b$.
\textbf{(B)} Uncertainty $\Delta B$ versus total interrogation time $T$.
Blue (green) dashed lines denote the numerical calculation of $\Delta B_{\rm FMM}$ for $T_R=T_1$ ($T_R=T_{\rm max}$) {according to the noise power spectral density} (see Eq. \ref{eq_sandard_deviation_ramsey_time}).
The red dash-dotted line is a fit of the Bayesian data with increasing $T_i$, showing sub-SQL scaling $\propto T^{-0.85\pm 0.01}$ and the red dashed line is the numerical calculation of $\Delta B_{\rm est}$ (see Eq.~\ref{accumulation_allan_Bayesian measurement_scaling}).
\textbf{(C)} Sensitivity $\eta=\Delta B\sqrt{T}$ (pT$/\sqrt{\rm Hz}$) versus total interrogation time $T$. The numerical calculations are plotted with dashed lines for frequentist measurement according to the noise power spectral density (see Eq.~\ref{eq_sensitivity_total_phase_accumulation_time}) and Bayesian measurement (see Eq.~\ref{accumulation_eta_Bayesian measurement_scaling}), respectively.
{The red dash-dotted line is a fit of the Bayesian data with increasing $T_i$, showing sub-SQL scaling $\propto T^{-0.35\pm 0.01}$.}
The error bars represent performance within 68.3$\%$ percentile range.
}
\label{Figure2}
\end{figure*}

Generally, a Bayesian quantum estimation procedure consists of a sequence of quantum interferometry of varying interrogation phases or interrogation times.
In our experiment, we implement a sequence of CPT-Ramsey interferometry that exponentially increases the interrogation time $T_R$ and adaptively updates the auxiliary phase $\phi_c$.
The schematic of our Bayesian cold-atom CPT magnetometry is shown in Fig. \ref{Figure2}A.
For convenience, we denote the interrogation time,  effective population number, and auxiliary phase in the $i$-th Bayesian update as $T_R^{(i)} \equiv T_i$, $N_{\rm eff}^{(i)}\equiv N_i=A e^{-2(T_i/T_{\chi})^2}$, and $\phi_c^i$.
Since there is no prior knowledge at the beginning, our Bayesian iterations start with a uniform prior distribution given by $p_1(B) = |\Delta m_F\gamma| T_1$.
After each phase-domain CPT-Ramsey interferometry, the conditional probability distribution $p_i(B|p_e^i,\phi_c^i,T_i,N_i)$ is updated according to the Bayes rule: $p_i(B|p_e^i,\phi_c^i,T_i,N_i) =\mathcal{N}\mathcal{L}_i(p_e^i|B,\phi_c^i,T_i,N_i)p_{i-1}(B\vert p_e^{i-1},\phi_c^{i-1},T_{i-1},N_{i-1})$~\cite{Degen_2017,Gebhart_2023}, where $\mathcal{N}$ is a normalization factor and $\mathcal{L}_i(p_e^i|B,\phi_c^i,T_i,N_i)$ is the likelihood function that gives the probability of the atoms occupying the state $\ket{F=2,m_F=-1}$ for a given $B$. 
The next update is implemented by inheriting the current posterior distribution as the next prior distribution $p_i(B) = p_{i-1}(B\vert p_e^{i-1},\phi_c^{i-1},T_{i-1},N_{i-1})$.
The auxiliary phase $\phi_c^i$ in the $i$-th iteration is determined by the previous posterior distribution {$p_{i-1}(B|p_e^{i-1},\phi_c^{i-1},T_{i-1},N_{i-1})$} at each iteration accordingly.
Finally, the estimated value after $M_b$ iterations is given by the mean $B_{\rm est} = \int Bp_{M_b}(B|p_e^{M_b},\phi_c^{M_b},T_{M_b},N_{M_b})dB$
over the posterior distribution, with uncertainty $\Delta B_{\rm est} = \sqrt{\int B^2p_{M_b}(B|p_e^{M_b},\phi_c^{M_b},T_{M_b},N_{M_b})dB-(B_{\rm est})^2}$.

In order to achieve high sensitivity in a wide dynamic range, the interrogation times exponentially increase according to $T_i=T_{\rm max}/a^{j-i}$ ($1\le i \le j$) before $T_i$ reaches $T_{\rm max}$ and then are fixed as $T_i=T_{\rm max}$ ($j< i \le M_b$).
Here, $a>1$ and $j = \log_a(T_{\rm max}/T_{\rm min}) + 1$.
In our Bayesian estimation procedure, the dynamic range is determined by the minimum interrogation time $T_{\rm min}$.
In our experiment, the available minimum interrogation time can be taken as \( T_{\rm min} \geq 0.2 \) ms, which corresponds to a dynamic range $B_{\rm max} \le$ 0.15 $\mu$T.

In addition to a sequence of correlated interferometry with varying interrogation times, a crucial aspect of our adaptive estimation procedure is the selection of optimal auxiliary phase $\phi_c^i$ for each interferometry. 
This selection is determined by previous measurements, allowing for a reduction in uncertainty when estimating the magnetic field \cite{Lumino_2018}.
To give $\phi_c^i$, we use the expected gain in Shannon information of the posterior distribution \cite{Ruster_2017},
\begin{equation}
    {U_{\phi_c}^i=\int^1_0dp_e\int U^i_{p_e,\phi_c}\mathcal{L}_i(p_e|B,\phi_c,T_i,N_i)p_i(B)dB,}
    \label{ee:utilize}
\end{equation}
where {$U_{p_e,\phi_c}^i = \int p_i(B|p_e,\phi_c,T_i,N_i) \ln [p_i(B|p_e,\phi_c,T_i,N_i)]dB - \int p_i(B) \ln [p_i(B)]dB$} denotes the expected gain in Shannon information of the posterior distribution with respect to the prior function after a hypothetical measurement, and {$\mathcal{L}_i(p_e|B,\phi_c,T_i,N_i)$} is the likelihood function. 
The ideal auxiliary phase for an upcoming measurement is one that maximizes the expected gain in Shannon information, i.e., $\phi_c^i=\arg \max_{\phi_c}[U_{\phi_c}^i]$.
Given the known result of the prior distribution, we introduce the auxiliary phase to ensure that the measurement slope is consistently close to its maximum. 
This approach minimizes the uncertainty associated with each individual measurement.

Compared to the frequentist scheme, our Bayesian scheme improves the scaling of sensitivity versus total interrogation time $T=\sum_{i=1}^{M_b}T_i$ to a sub-SQL scaling.
Attribute to the Bayesian update, the uncertainty can be given as $\Delta B_{\rm est}\approx C/\sqrt{\sum_{i=1}^{M_b}N_iT_i^2}$ with $C=1/(2\pi|\Delta m_F\gamma|)$ (see details in Section 4 of Supplementary Material).
The uncertainty follows a sub-SQL scaling $\Delta B_{\rm est} \propto T^{-0.85\pm0.01}$ when $i < j$, and gradually converges to the SQL scaling $\Delta B_{\rm est} \propto T^{-0.5}$ when $i \gg j$, see Fig.~\ref{Figure2}B.
When $i \gg j$, the SQL scaling of our Bayesian scheme can be analytically given as
\begin{equation}
    {\Delta B_{\rm est}\approx
    \frac{C}{\sqrt{N_j T_{\rm max} T}}\propto T^{-0.5},\ i \gg j.}
    \label{accumulation_allan_Bayesian measurement_scaling}
\end{equation}
Consequently, the sensitivity with respect to the total interrogation time $T$ is given as
\begin{equation}
    {\eta_{\rm est}=\Delta B_{\rm est} \sqrt{T}\approx
    \frac{C}{\sqrt{N_j T_{\rm max}}},\ i \gg j .}
    \label{accumulation_eta_Bayesian measurement_scaling}
\end{equation}
The sensitivity follows a scaling  $\eta_{\rm est} \propto T^{-0.35\pm0.01}$ when $i < j$, and converges to a fixed value when $i \gg j$, see Fig.~\ref{Figure2}C.

\subsection*{Sensitivity and dynamic range}

\begin{figure}[htp]
	\includegraphics[width=1\linewidth]{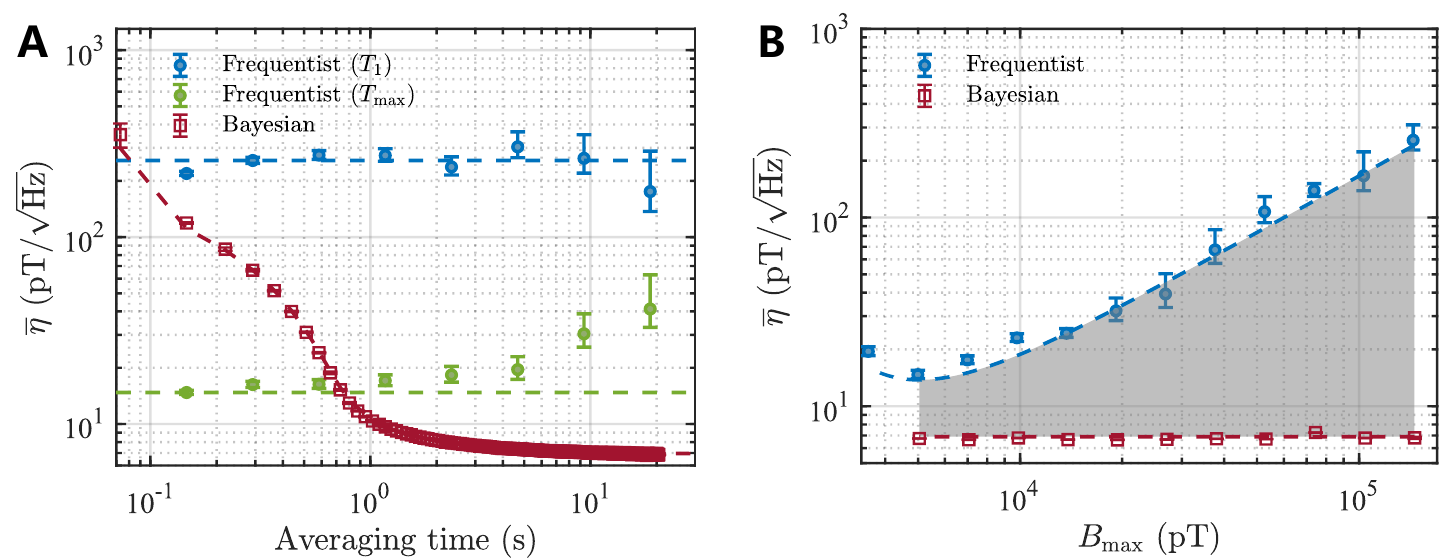}
	\caption{
\textbf{Sensitivity and dynamic range.}
\textbf{(A)} Sensitivity $\overline{\eta}=\Delta B \sqrt{\tau}$ for frequentist measurements (blue and green circles) and Bayesian measurements (red squares).
The averaging time $\tau$ includes interrogation time and dead time.
The blue and green dashed lines denote the results from noise power spectral density for frequentist measurements with $T_R=T_1$ and $T_R=T_{\rm max}$, respectively.
The red dashed line corresponds to the numerical results for Bayesian measurements, which gradually converges to a fixed value $C{\sqrt{T_c}}/{(\sqrt{N_j}T_{\text{max}})}$.
\textbf{(B)} Sensitivity $\overline{\eta}=\Delta B \sqrt{\tau}$ versus dynamic range $B_{\rm max}$. 
The gray shaded area indicates the gain in dynamic range and sensitivity of Bayesian measurements compared to frequentist measurements.
{The blue dashed curve represents the numerical results for frequentist measurements with different $T_R$ according to Eq. (\ref{eq_sensitivity_averaging_time}). 
}
}
\label{Figure3}
\end{figure}

In a Bayesian quantum estimation, the dynamic range preserves the highest value imposed by the first interferometry of the minimum interrogation time, while the sensitivity is gradually improved via Bayesian updates.
As a sensor always has a dead time, the sensitivity with respect to the averaging time $\tau$ can truly reflect its performance.
In our Bayesian quantum magnetometry, the sensitivity versus the averaging time $\tau=M_b T_c$ obeys $\overline{\eta}_{\rm est}=\Delta B_{\rm est} \sqrt{\tau}$ and the dynamic range is given as \(B_{\rm max} = 1/(2 |\Delta m_F \gamma| T_1)\).
We experimentally demonstrate how to improve the sensitivity and dynamic range of our cold-atom CPT magnetometer via Bayesian quantum estimation.
For comparison, the highest dynamic range ($T_R=T_1$) and the highest sensitivity ($T_R=T_{\rm max}$) associated with frequentist measurements are also presented (see Fig. \ref{Figure3}).

In frequentist measurements, the sensitivity {$\overline{\eta}_{\rm FMM} \propto B_{\rm max}/\sqrt{N_{\rm eff}}$} becomes worse when the dynamic range $B_{\rm max}$ increases.
The best sensitivity of 14.7$\pm$0.4~pT$/\sqrt{\rm Hz}$ at an averaging time of 0.146~s is achieved with the interrogation time $T_R=T_{\rm max}=7.1$~ms, corresponding to the lowest dynamic range of 5.0~nT.
The highest dynamic range of 145.6~nT is obtained with the minimum interrogation time $T_R=T_1=0.245$~ms, corresponding to the worst sensitivity of 256.8$\pm$10.1~pT$/\sqrt{\rm Hz}$ at an averaging time of 0.292~s.
As the effective particle number $N_{\rm eff}$ is influenced by the SNR which decreases with $T_R$, the sensitivity $\overline{\eta}_{\rm FMM}$ is not a linear function of $B_{\rm max}$, but exhibits an optimal point (see Fig. \ref{Figure3}B).

In Bayesian measurements, the dynamic range is determined by the first interrogation time, which is also the minimum interrogation time in the interferometry sequence.
By choosing $a=1.4$, $T_{\rm max}=7.1$~ms and the first interrogation time $T_1=0.245$~ms, the corresponding dynamic range is 145.6~nT.
Meanwhile, when \( i \gg j \), the sensitivity \(\overline{\eta}_{\rm est}\) gradually converges to a fixed value \(C{\sqrt{T_c}}/{(\sqrt{N_j}T_{\text{max}})}\) (see details in Section 5 of Supplementary Materials).
The Bayesian scheme achieves a sensitivity of 6.8$\pm$0.1~pT$/\sqrt{\rm Hz}$ at an averaging time of $\tau=18.031$~s by $M_b=247$ iterations, see Fig.~\ref{Figure3}~A.
For frequentist measurements taken with $T_R = T_{1}$, the dynamic range is the same and the optimal sensitivity {256.8$\pm$10.1} pT$/\sqrt{\rm Hz}$ is achieved at an averaging time of {$\tau=0.292$}~s, our Bayesian scheme gives a {15.8$\pm$0.2} dB enhancement in sensitivity; see Table \ref{tab:my_label}.
For the frequentist measurement taken with $T_R = T_{\rm max}$, the optimal sensitivity 14.7$\pm$0.4~pT$/\sqrt{\rm Hz}$ is achieved at an averaging time of $\tau=0.146$~s, our Bayesian scheme still has an enhancement of 3.3$\pm$0.1~dB in sensitivity, while the dynamic range is improved by 14.6~dB, see Table~\ref{tab:my_label}.
The sensitivity gain $Q=\overline{\eta}_{\rm FMM}/\overline{\eta}_{\rm est}=\frac{2T_{\rm max}\sqrt{N_j}}{T_R\sqrt{N_{\rm eff}}}$ comes from two aspects.
On the one hand, $T_R\sqrt{N_{\rm eff}}$ increases monotonically between $T_1$ and $T_{\rm max}$, until reaching its maximum value at $T_{\rm max}$. 
Consequently, compared to the frequentist measurements taken with $N_{\rm eff} = N_1$ and $T_R=T_1$ (see Eq.~\ref{eq_sensitivity_averaging_time}), the optimal gain of $Q=2T_{\rm max}\sqrt{N_j}/(T_1\sqrt{N_1})=15.4$~ dB is achieved.
On the other hand, the determination of $f_p$ requires two individual CPT-Ramsey interferometry in frequentist measurements; therefore, Bayesian measurements still yield a two-fold improvement in sensitivity, i.e., $Q=$ 3 dB, over the frequentist measurements taken with $N_{\rm eff} = N_j$ and $T_R = T_{\rm max}$ (see Eq.~\ref{eq_sensitivity_averaging_time}).
We experimentally demonstrate that the sensitivity is independent of the dynamic range, by choosing different $T_1$ to perform Bayesian measurements (see Fig. \ref{Figure3}B).

\begin{table}
\renewcommand{\arraystretch}{1.5}
    \centering
    \caption{\textbf{Comparison between frequentist and Bayesian measurements.}
    {The achievable sensitivities and dynamic ranges are presented for both frequentist and Bayesian measurements. 
    The sensitivities of ideal frequentist measurements should be independent of the averaging time, see blue and green dashed lines in Fig. \ref{Figure3}A.
    In realistic experiments, due to low-frequency noises, the sensitivity may deteriorate when the averaging time increases.
    Therefore, the optimal sensitivity for frequentist measurements is obtained at short averaging times, where the sensitivities are also coincident with the noise power spectral density.
    Meanwhile, long averaging time allows the sensitivity of Bayesian measurements to approach the theoretical limit $C\sqrt{T_c}/(\sqrt{N_j}T_{\rm max})$.
    }
    }
    \begin{tabular}{cccc}
    \hline
        ~ & optimal sensitivity $\overline{\eta}$ & dynamic range $B_{\rm max}$\\
        ~ & (pT$/\sqrt{\rm Hz}$) & (nT)\\
        \hline 
        Frequentist measurements with $T_1$ & {256.8$\pm$10.1} & {145.6}\\
        Frequentist measurements with $T_{\rm max}$ & {14.7$\pm$0.4} & {5.0}\\
        Bayesian measurements & {6.8$\pm$0.1} & {145.6}\\
        \hline 
    \end{tabular}
    \label{tab:my_label}
\end{table}

\subsection*{Magnetic-field tracking}

\begin{figure}[ht]
	\centering\includegraphics[width=0.5\linewidth]{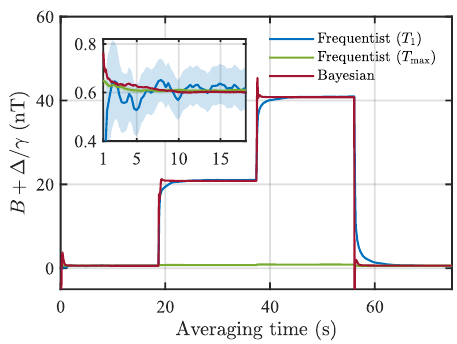}
	\caption{
\textbf{Magnetic-field tracking.}
Magnetic-field tracking when the value of $B$ is stepped by 20~nT at 18.031~s intervals.
The solid red line represents typical performances of Bayesian measurements with $a=1.4, M_b = 247$.
For comparison, two frequentist measurements with interrogation time $T_R=T_1$ (blue line) and $T_R=T_{\rm max}$ (green line) are presented. 
The shaded area indicates the uncertainty. 
The inset shows the enlarged region of magnetic field tracking from 1~s to 18.031~s to compare the uncertainties of three different tracking curves. 
}
\label{Figure4}
\end{figure}

In realistic systems, the magnetic field may vary with time.
To verify the tracking capability of time-varying magnetic fields, we increase the field strength by 20~nT at 18.031~s intervals, performing two increments in total. 
This was followed by a restoration of the magnetic field with a step change of 40~nT.
For frequentist measurements, the dynamic range and sensitivity are inversely proportional.
Higher sensitivity means a smaller detectable magnetic field change.
Therefore, the frequentist measurements taken with $T_R=T_{\rm max}$, which has a dynamic range of 5.0~nT, cannot respond to the change of 20~nT (see the green line in Fig.~\ref{Figure4}).
If the interrogation time is fixed as $T_R=T_1$, the frequentist measurement can respond to the change of 20~nT, but has low sensitivity (see the blue shaded area in Fig.~\ref{Figure4}).

In contrast, Bayesian measurements operate with exponentially growing interrogation times, which ensure superior sensitivity while maintaining high dynamic range. 
We compare the Bayesian quantum magnetometry {($a=1.4,\ M_b = 247$)} with the frequentist measurements that achieve the maximum dynamic range using $T_R=T_1$ or the maximum sensitivity using $T_R=T_{\rm max}$. 
The estimated values for a static d.c. magnetic field are consistent with each other.
The uncertainty $\Delta B_{\rm est}$ obtained by Bayesian protocol is the smallest among the three protocols.
Furthermore, when we suddenly change $B$, the estimated values of the Bayesian protocol can converge to the corresponding value after approximately 20 iterations.
The experimental data clearly show that, in comparison to the conventional frequentist protocol, our Bayesian cold-atom CPT magnetometry has the ability to track time-varying magnetic fields with larger dynamic range while maintaining higher sensitivity.

\section*{Discussion}
\noindent We have experimentally demonstrated an adaptive high-precision measurement of the d.c. magnetic field with a CPT magnetometer of cold $^{87}$Rb atoms.
By implementing a sequence of correlated CPT-Ramsey interferometry guided by our algorithm, the measurement sensitivity achieves a sub-SQL scaling with respect to the total interrogation time as $\propto T^{-0.35\pm0.01}$.
We obtain a measurement sensitivity of \(6.8\pm0.1 \ \mathrm{pT}/\sqrt{\mathrm{Hz}}\) at an averaging time of $18.031~\textrm{s}$ with a dynamic range of {145.6~nT}.
Compared to the frequentist measurement taken with the longest individual interrogation time $T_R=T_{\rm max}$, which gives a sensitivity of \(14.7\pm 0.4 \ \mathrm{pT}/\sqrt{\mathrm{Hz}}\) and a dynamic range of 5.0 nT, our results represent an improvement of $3.3\pm0.1$ dB in sensitivity and 14.6 dB in dynamic range.
Our study opens avenues for the next generation of adaptive cold-atom quantum sensors, wherein real-time measurement history is leveraged to improve their performance.

In contrast to conventional cold-atom magnetometry, which offers high sensitivity but limited dynamic range~\cite{Sekiguchi_2021}, our adaptive cold-atom CPT magnetometry not only maintains superior sensitivity and high spatial resolution, but also achieves a significantly improved dynamic range.
In our experiments, the sensitivity and spatial resolution are constrained by the free-fall motion of the atoms and the decoherence during the CPT-Ramsey interference process. 
On the one hand, using atoms trapped in optical traps to facilitate in situ CPT-Ramsey interference would improve the spatial resolution.
On the other hand, trapping atoms at magic wavelength and magic intensity~\cite{Lundblad_2010, Guo_2020, Li_2019} could extend their coherence time, further improving the sensitivity.
If the longest interrogation time of our adaptive cold-atom CPT magnetometry is extended to 300~ms, the sensitivity can be improved to 513~fT$/\sqrt{\rm Hz}$ without compromising the dynamic range.

\section*{Materials and Methods}

\subsection*{Evaluation of the coherence time}
At beginning, the atoms are prepared into the dark state by applying a CPT pulse. 
The state evolves over time $T_R$ and the CPT-Ramsey fringe could be obtained by detecting the transmission signal during another CPT pulse.
When the excited-state decay rate $\Gamma$ is large compared to all other decay rates, we can apply an adiabatic approximation to the time-evolution of the excited-state $\ket{3}$ based on a three-level CPT system.
The transmitted signal $S_{TS}$ containing the ground-state coherence is given by the expression, 
\begin{equation}
    S_{TS} \propto \int \left[1-\rho_{33}(\textbf{r})\right] d\textbf{r},
\end{equation}
where $\rho_{33}(\textbf{r})$ is the population in the excited state $\ket{3}$ and it can be written as  \cite{Hemmer_1989,Pati_2015}
\begin{equation}
    \rho_{33}(\textbf{r})=\alpha(\textbf{r}) e^{-\alpha(\textbf{r})\Gamma\tau_d}\left\{1-\left[1-e^{-\alpha(\textbf{r})\Gamma\tau_p}\right]|\sec{(\phi_{ls})}|\cos{\left[2\pi(\Delta f- f_B)T_R-\phi_{ls}\right]}\right\}.
    \label{equation_rho33}
\end{equation}
Here, $\alpha=\Omega(\textbf{r})^2/(\Gamma^2+3\Omega(\textbf{r})^2+4\Delta f^2)$, $\Omega(\textbf{r})$ is the average Rabi frequency, $f_B = \Delta m_F\gamma B$ is the Larmor frequency, and $\phi_{ls}$ denotes the phase shift.
The phase shift $\phi_{ls}$ can be ignored when $\tau_p$ completely prepares the atoms into the dark state \cite{Pati_2015}. 
According to Eq.~(\ref{equation_rho33}), time-domain CPT-Ramsey interference is obtained by scanning the interrogation time $T_R$.
In cold-atom CPT-Ramsey interferometry, the amplitude of time-domain CPT-Ramsey interference varies due to the Rabi frequencies $\Omega(\textbf{r})$ changes when the atomic cloud falls due to the gravity.
By assuming the laser intensity varies {parabolically} in the short distance that the atomic cloud crosses {the center of the} laser beam, the Rabi frequency versus the position $z$ can be given as,
\begin{equation}
    {{\Omega(\textbf{r})\approx\Omega_0(1-kz^2),}}
\end{equation}
where $\Omega_0$ = 0.18 MHz is the initial average Rabi frequency (the atomic cloud is initially positioned {at the center} of the CPT light), {$k={1}/(2\sigma_{c}^2)$ is the second-order coefficients} in the Taylor expansion at the initial position, the width of CPT light {$\sigma_c=1.83$} mm, and $z=1/2gT_R^2$ with the gravity acceleration $g$. 
Assuming that the average Rabi frequency is independent upon $x$ and $y$, this expression keeps valid for the {17}-ms free fall corresponding to {$z=1.4$} mm.
Taking decoherence into account, the transmitted signal obeys 
\begin{equation}
    S_{TS} \propto 1-\alpha(z) e^{-\alpha(z)\Gamma\tau_d}\left\{1-\left[1-e^{-\alpha(z)\Gamma\tau_p}\right]e^{-(T_R/T_\chi)^2}\cos{\left[2\pi(\Delta f- f_B)T_R\right]}\right\}.
\end{equation}
According to Eq. (10), we fit the Ramsey fringes and obtain the coherence time {$T_\chi=10.0$} ms.
\\

\subsection*{Determination of the effective particle number}
Due to the decoherence and $z$-dependent Rabi frequency, the amplitude of time-domain CPT-Ramsey fringes decreases with the interrogation time $T_R$.
In frequency-domain and phase-domain CPT-Ramsey interferometry, the interrogation time $T_R$ is fixed and so that the influence of decoherence and $z$-dependent Rabi frequency is transformed into the change of SNR. 
Meanwhile, in both frequentist and Bayesian measurements, it is more convenient to use the normalized signal
\begin{equation}
    p_e=\frac{1}{2}\left\{1-\cos{\left[2\pi(\Delta f- f_B)T_R+\phi_c\right]}\right\},
\end{equation}
instead of {$S_{TS}$}.

In our experiments, the SNR can be defined as \cite{Degen_2017}
\begin{equation}
        {\rm SNR}=\frac{\delta p_{\rm obs}}{\sigma_p}=\delta p e^{-\chi(T_R)}2D\sqrt{M}\sqrt{N},
\end{equation}
where $\delta p_{\rm obs}=\delta p e^{-\chi(T_R)}$ is the reduction of the observed probability, $\sigma_p^2=1/(4D^2MN)$ is the total readout uncertainty, $\chi(T_R)=(T_R/T_\chi)^2$ is the phenomenological decoherence function, $D\le1$ describes the reduction of the signal-to-noise ratio compared to an ideal readout ($D$ = 1), $M$ is the number of measurements, and $N$ denotes the total population number. 
To further specify the SNR, the change in probability $\delta p$ is related to the change in signal $\delta B$ as $\delta p=\delta B |\partial p(T_R)/\partial B|=$$\pi|\Delta m_F\gamma| T_R \delta B$ for slope detection.
In conventional frequentist measurements, two CPT-Ramsey cycles are required to complete one measurement, thus the uncertainty $\sigma_p^2$ doubly increases.
For a given averaging time $\tau$, the number of measurements is $M=\tau/[2(T_R+T_d)]$ with $T_d$ being the extra dead time needed to prepare, initialize, and read out.
Thus, the SNR can be given as
\begin{equation}
        {\rm SNR}=\pi|\Delta m_F \gamma| T_R \delta B e^{-\chi(T_R)}D\sqrt{N}\frac{\sqrt{\tau}}{\sqrt{T_R+T_d}}.
\end{equation}

The sensitivity is defined as the minimum detectable signal that yields unit SNR for an averaging time of one second,
\begin{equation}
        \overline{\eta}_{\rm FMM}=\delta B_{\rm min}=\frac{\sqrt{T_R+T_d}}{\pi|\Delta m_F \gamma| T_R e^{-\chi(T_R)}D\sqrt{N}}.
\end{equation}
Defining the effective number $N_{\rm eff}\equiv \left[e^{-\chi(T_R)}D\sqrt{N}\right]^2$, we have the sensitivity  
\begin{equation}\label{sensitivity}
    \overline{\eta}_{\rm FMM}= \frac{\sqrt{T_R+T_d}}{\pi |\Delta m_F \gamma| T_R \sqrt{N_{\rm eff}}}.
\end{equation} 

In experiments, one can determine the effective particle number $N_{\rm eff}=\frac{T_R+T_d}{(\pi |\Delta m_F \gamma| T_R \overline{\eta}_{\rm FMM}^{\rm exp})^2}$ from the sensitivity $\overline{\eta}_{\rm FMM}^{\rm exp}$ derived from the noise power spectral density.
In our experiments, $N_{\rm eff}$ approximately decreases as $N_{\rm eff}=(10755\pm2101) e^{-2(T_R/T_{\chi})^2}$, and optimal integration time corresponding to the best sensitivity is $7.1$ ms (see Supplementary Material Section 2).

\subsection*{Basic procedure of Bayesian cold-atom CPT magnetometry}
In a single-particle Ramsey interferometry, the likelihood function reads $\mathcal{L}_u(u|B,\phi_c{, T_R})\ =\ \frac{1}{2}\{1+(-1)^{u}\cos{[2\pi(\Delta f-\Delta m_F\gamma B){T_R}+\phi_c]}\}$, where $u=$ 0 or 1 stands for the particle occupying the magnetic sensitive state $\ket{F=1,m_F=-1}$ or $\ket{F=2,m_F=-1}$ respectively.
In CPT-Ramsey interferometry, the signal of each measurement is provided by an ensemble of atoms rather than a single atom. 
This means that the probability $p_e$ of the atoms occupying the magnetic sensitive state $\ket{F=2,m_F=-1}$ obeys a binomial distribution, which can be approximated by a Gaussian distribution when the total particle number is sufficiently large. 
Below we use a Gaussian distribution function as our likelihood function,
\begin{equation}
    {\mathcal{L}(p_e|B,\phi_c,T_R,N_{\rm eff})\ =\ \frac{1}{\sqrt{2\pi}\sigma}\exp{\left\{-\frac{\left[p_e-\mathcal{L}_u(1|B,\phi_c,T_R)\right]^2}{2\sigma^2}\right\}},}
    \label{eqBayesian measurement:mp_Likelihood}
\end{equation}
where {$\sigma^2\approx p_e(1-p_e)/N_{\rm eff}$, $N_{\rm eff}=A e^{-2(T_R/T_\chi)^2}$ is effective particle number, and $A$ is a constant.}

The initial prior distribution of $B$ is set as a uniform distribution over the interval $[B_l, B_r]$ of a width
$B_{lr} \equiv B_r - B_l = 1/(|\Delta m_F \gamma| T_1)$. 
To implement our magnetometry protocol, the interval $[B_l, B_r]$ should include the value $B$ to be estimated.
From the prior function $p_i(B)$, the posterior function in the $i$-th Bayesian update is calculated through the Bayes’ formula, 
\begin{equation}
    {p_i(B|p_e^i, \phi_c^i, T_i, N_i) = \mathcal{N}\mathcal{L}_i(p_e^i|B,\phi_c^i,T_i,N_i)p_{i}(B),}
    \label{eqBayesian measurement:Bayes_formular}
\end{equation}
where $\mathcal{N}$ is a normalization factor.
An estimation of $B$ and its uncertainty can be given as {$B^{(i)}_{\rm est}=\int B p_i(B|p_e^i, \phi_c^i, T_i, N_i) dB$ and $\Delta B^{(i)}_{\rm est}=\sqrt{\int B^2p_i(B|p_e^i, \phi_c^i, T_i, N_i)dB-(B_{\rm est}^{(i)})^2}$, respectively.}
The next update is implemented by inheriting the posterior function as the next prior function, that is, $p_{i+1}(B)=p_i(B|p_e^i, \phi_c^i, T_i, N_i)$.
Given the $i$-th interrogation time $T_i$, the corresponding interval is turned into $[B^{(i)}_{\rm est} - 1/(2|\Delta m_F\gamma| T_i), B^{(i)}_{\rm est} + 1/(2|\Delta m_F\gamma| T_i)]$.
Subsequently, the previous distribution $p_i(B)$ is reset according to the estimated value $B_{\rm est}^{(i)}$ and the estimated uncertainty $\Delta B_{\rm est}^{(i)}$ given by the previous step.

In the adaptive procedure, the auxiliary phase $\phi_c^i$ for the $i$-th update is determined by the previous posterior distribution $p_{i-1}(B|p_e^{i-1}, \phi_c^{i-1}, T_{i-1}, N_{i-1})$.
To give $\phi_c^i$, we use the expected gain in Shannon information of the previous posterior distribution, which is expressed by the Utilize function $U_{\phi_c}^i$ (Eq. \ref{ee:utilize}).
Here, we calculate the Utilize function by discretizing the integral over $p_e$ (see Supplementary Material Section 6). 
Thus, $\phi_c^i$ is chosen as the one that maximizes $U_{\phi_c}^i$.
Once \( \phi_c^i \) and \( T_i \) are given, measurements can be conducted to obtain the probability $p_e^i$.
After $M_b$ iterations, we reset the prior to the initial distribution $p_1(B) = |\Delta m_F \gamma| T_1$ to accommodate typical sensing experiments where the strength of $B$ is not fixed.

The basic workflow of our Bayesian cold-atom CPT magnetometry is implemented according to the following flowchart.
\begin{itemize}
\item Step 1: Determine the values of all input parameters $\{T_{\rm min},T_{\rm max},{T_\chi,A},a,M_b\}$.
\item Step 2: Initialize the magnetic field interval $[B_l, B_r]$ and the prior distribution $p_1(B)$. 
The interval should include the value $B$ to be estimated (i.e. $B_l<B<B_r$) and the interval width $B_{lr} \equiv B_r-B_l=1/(|\Delta m_F\gamma| T_1)$. 
The initial prior distribution is chosen as the uniform distribution over the interval $[B_l, B_r]$, i.e. $p_1(B)=|\Delta m_F \gamma| T_1$.
\item Step 3: Implement the loop. 
(a) The interrogation time $T_i$ is given by $T_i=T_{\rm max}/a^{j-i}$ if $i < j$, and $T_i=T_{\rm max}$ if $i \ge j$. 
Here,  $a>1$, $i=1,2,...,M_b$, $j$ = $\log_a(T_{\rm max}/T_{\rm min})$ + 1 is the number that interrogation time $T_i$ increases from $T_{\rm min}$ to $T_{\rm max}$.
{The effective particle number $N_i=Ae^{-2(T_i/T_\chi)^2}$.}
The interval width $B_{lr}$ is updated according to $\left[B^{(i)}_{\rm est} - 1/(2|\Delta m_F\gamma| T_i), B^{(i)}_{\rm est} + 1/(2|\Delta m_F\gamma| T_i)\right]$. 
The prior distribution $p_i(B)$ is reset according to the estimated value $B_{\rm est}^{(i)}$ and the estimated uncertainty $\Delta B_{\rm est}^{(i)}$ given by the previous step.
The auxiliary phase $\phi_c^i$ is obtained by maximizing the Utilize function of Eq. \ref{ee:utilize}.
(b) Conduct experiment to obtain the population probability $p_e$ with $T_i$ and $\phi_c^i$.
(c) Perform Bayesian iteration. The likelihood function is defined by Eq.~\protect\ref{eqBayesian measurement:mp_Likelihood}. 
The probability distribution is updated as a posterior distribution $p_i(B|p_e^i,\phi_c^i.T_i,N_i)$ according to Bayes' formula of Eq. \ref{eqBayesian measurement:Bayes_formular}. 
The estimated value and uncertainty can be obtained from the posterior distribution. 
(d) The next update is implemented by inheriting the posterior distribution as the next prior distribution. 
\item Step 4: After $M_b$ iterations, we reset the prior distribution as the initial distribution $p_1(B) = |\Delta m_F \gamma| T_1$.
\end{itemize}

Repeat execution of steps 2 to 4 for the next measurement.

The algorithm used in our experiment is shown in Algorithm \ref{BPE_algorithm}.

\begin{algorithm}
\caption{Flowing chart of adaptive Bayesian cold-atom CPT magnetometry}
\label{BPE_algorithm}
\SetKwInOut{Input}{Input}
\SetKwInOut{Output}{Output}
\SetKwInOut{Initialize}{Initialize}
\SetKwFunction{Actor}{Actor}
\SetKwFunction{Critic}{Critic}
\SetKwFunction{Softmax}{Softmax}
\SetKwFunction{CrossEntropy}{CrossEntropy}

\BlankLine
\Input{parameters $\{T_{\rm min},T_{\rm max},T_\chi,A,a,M_b\}$.}
\Initialize{initial interval $[B_l, B_r]$; initial prior distribution $p_1(B) = |\Delta m_F\gamma| T_1$; number of different interrogation times $j$ = $\log_a(T_{\rm max}/T_{\rm min})$ + 1.}
\BlankLine
[Bayesian magnetic field measurement Loop]:\\
\For{$i = 1$ \KwTo $M_{b}$}{
\BlankLine
[Updates of parameter]\;
interrogation time: $T_i=\left\{
    \begin{array}{ll}
	T_{\rm max}/ a^{j-i}, & i < j\\
    T_{\rm max}, &  i \ge j
    \end{array} \right. $  \;
    effective particle number: $N_i=A e^{-2(T_i/T_\chi)^2}$\;
      length of interval: $B_{lr} = 1/(|\Delta m_F \gamma| T_i$)\;
       \If{$i>1$}{
          $B_l \leftarrow B_{\rm est}^{(i)}-B_{lr}/2$\;
          $B_r \leftarrow B_{\rm est}^{(i)}+B_{lr}/2$\;
          reset the prior distribution $p_i(B) = \frac{1}{\sqrt{2\pi}{\sigma^i_B}} \exp{[- \frac{(B-\mu_i)^2}{2{(\sigma^i_B)}^2}] }$, where $\mu_i=B_{\rm est}^{(i)}$ and ${\sigma^i_B} = \Delta B^{(i)}_{\rm est}$
       }
      adaptive update of auxiliary phase: $\phi_c^{i} = \arg{\max_{\phi_c}{{U_{\phi_c}^i}}}$
      \BlankLine
      [Experimental measurement]\;
      measuring probability $p_e$ using $T_i$ and $\phi_c^i$\;
      \BlankLine
      [Bayesian iteration]\;
      Likelihood function:    
        {$\mathcal{L}_i(p_e^i\vert B, \phi_c^i,T_i,N_i)= \frac{1}{\sqrt{2\pi}\sigma_i}\exp{\left[-\frac{(p_e^i-\mathcal{L}_u(1\vert B,\phi_c^i,T_i))^2}{2\sigma_i^2}\right]}$},
      where {$\sigma_i^2 = p_e^i(1-p_e^i)/N_i$}\;
      Bayesian update: {$p_i(B|p_e^i,\phi_c^i,T_i,N_i) \leftarrow \mathcal{N} \mathcal{L}_i(p_e^i|B, \phi_c^i,T_i,N_1) p_i(B)$}\;
      Estimated magnetic field: {$B_{\rm est}^{(i)} = \int{B p_i(B|p_e^i,\phi_c^i,T_i,N_i) \mathrm{d} B}$}\;
      Uncertainty:
      {$\Delta B_{\rm est}^{(i)} =
        \sqrt{\int{B^2 p_i(B|p_e^i,\phi_c^i,T_i,N_i) \mathrm{d} B} - (B_{\rm est}^{(i)})^2}$}\;
      Sensitivity with respect to total interrogation time:
      $\eta^{(i)}_{\rm est}=\Delta B_{\rm est}^{(i)}\sqrt{T}$, with $T=\sum_i T_i$\;
      Sensitivity with respect to averaging time:
      $\overline{\eta}^{(i)}_{\rm est}=\Delta B_{\rm est}^{(i)}\sqrt{\tau}$, with $\tau=i \cdot T_c$ \;
      \Output{estimated magnetic field $B_{\rm est}^{(i)}$; uncertainty $\Delta B_{\rm est}^{(i)}$; sensitivity $\eta^{(i)}_{\rm est}$ with respect to $T$; sensitivity $\overline{\eta}^{(i)}_{\rm est}$ with respect to $\tau$.}
    }
    [Reset]: reset the prior distribution to $p_1(B) = |\Delta m_F\gamma| T_1$. 
\BlankLine
\end{algorithm}

 \newpage
  

\section*{Acknowledgments}
\textbf{Funding:} This work is supported by the National Natural Science Foundation of China (grants 12025509 and 92476201 to C.L., grant 12104521 to C.H., and grant 12475029 to J.H.); the China National Key Research and Development Program (grant 2022YFA1404104 to B.L., C.L. and C.H.); and Guangdong Provincial Quantum Science Strategic Initiative (grants GDZX2305006 and GDZX2405002 to C.L., and grant GDZX2405003 to C.H.).\\
\textbf{Author contributions:} C.L., C.H., and B.L. conceived the project. Z.M., C.H., J.H., and C.L. developed the physical protocol. C.H., Z.M., and B.L. designed the experiment. C.H., Z.M., Z.T., H.H., S.S, X.K., J.W., and B.L. performed the experiment. All authors discussed the results and contributed to compose and revise the manuscript. C.L. supervised the project.\\
\textbf{Competing interests:} The authors declare no competing interests.\\
\textbf{Data and materials availability:} All data needed to evaluate the conclusions in the paper are present in the paper and/or the Supplementary Materials

\FloatBarrier 
 
\section*{Supplementary materials}
\noindent \textbf{This PDF file includes:}\\
 Sections S1 to S6\\
 Figs. S1 to S4

\clearpage

\end{document}